\documentclass[sn-mathphys,iicol]{sn-jnl}

\usepackage{tabularx}
\usepackage{subfig}
\usepackage{float}
\usepackage{multirow}
\usepackage{comment}
\usepackage{amsmath}
\usepackage{fancyref}
\usepackage{booktabs}

\usepackage{array}
\usepackage{arydshln}
\makeatletter
\renewcommand*\env@matrix[1][\arraystretch]{%
  \edef\arraystretch{#1}%
  \hskip -\arraycolsep
  \let\@ifnextchar\new@ifnextchar
  \array{*\c@MaxMatrixCols c}}
\makeatother

\jyear{2023}%

\theoremstyle{thmstyleone}%
\theoremstyle{thmstyletwo}%

\theoremstyle{thmstylethree}%

\begin{document}

\title{Stock Trend Prediction: A Semantic Segmentation Approach}
%

\author[1]{\fnm{Shima} \sur{Nabiee}}\email{snabiee@uci.edu}

\author[1]{\fnm{Nader} \sur{Bagherzadeh}}\email{nader@uci.edu}

\affil[1]{\orgdiv{Department of Electrical Engineering and Computer Science}, \orgname{University of California in Irvine}, \orgaddress{\city{Irvine}, \postcode{92697}, \state{California}, \country{USA}}}


\abstract{Market financial forecasting is a trending area in deep learning. Deep learning models are capable of tackling the classic challenges in stock market data, such as its extremely complicated dynamics as well as long-term temporal correlation. To capture the temporal relationship among these time-series data, recurrent neural networks are employed. However, it is difficult for recurrent models to learn to keep track of long-term information.  Convolutional Neural Networks (CNN) have been utilized to better capture the dynamics and extract features for both short-term and long-term forecasting. However, semantic segmentation and its well-designed fully convolutional networks have never been studied for time-series dense classification. We present a novel approach to predict long-term daily stock price change trends with fully 2D-convolutional encoder-decoders. We generate input frames with daily open, high, low, and close prices for a time-frame of T days. The aim is to predict future trends by pixel-wise classification of the current price frame. We propose a hierarchical CNN structure to encode multiple price frames to multiscale latent representation in parallel using Atrous Spatial Pyramid Pooling (ASPP) blocks and take that temporal coarse feature stacks into account in the decoding stages. Our hierarchical structure of CNNs make it capable of capturing both long and short-term temporal relationships effectively. The effect of increasing the input time horizon via incrementing parallel encoders has been studied with interesting and substantial changes in the output semantic segmentation maps. We achieve overall accuracy and AUC of \%78.18 and 0.88 for joint trend prediction over the next 20 days, surpassing other semantic segmentation approaches. One variation of the proposed framework results in \%83.19 in accuracy for four prices trends predictions of the next day, and the other variation achieves \%88.14 for the tenth day, the highest in a 20-day time-frame. Finally, we compared our proposed model with several deep learning models that were specifically designed for technical analysis and found that for different output horizons, our proposed model variations outperformed the other models.}



\maketitle

\section{Introduction}
Stock price and trend forecasting remain a classic complex field in economics. Huge challenges exist due to the random, non-stationary, non-linear, and noisy behavior of stock markets, as well as highly interconnected influencing factors such as political, economic, media, and psychological variables. Over recent years, deep learning models have become predominant and powerful tools in financial market analysis through non-linear, multivariate, and data-driven analysis, with numerous promising results in the domain \cite{LONG2020106205,info12060250,anand2021comparison,mehtab2020time}. Among relevant variants in deep learning models, Deep Neural Networks (DNN), Convolutional Neural Networks (CNNs), Recurrent Neural Networks (RNN), and Long Short-Term Memory (LSTM), have been widely presented in stock trend forecasting along with other time series classification problems.\\

RNNs with LSTM have been extensively adopted as the main elements for sequence analysis \cite{cho2014learning,chung2014empirical}. These models generate sequential hidden states, each depending on the current timestep’s input features and the previous hidden state. While recurrent models are conceptually intriguing for time series analysis, it has been found that for many tasks across domains, feed-forward models can substitute for recurrent models without sacrificing accuracy \cite{chen2017cnn,bai2018empirical}. Besides, their inherently sequential procedure precludes parallelization and can become the performance bottleneck. \cite{park2016fully} demonstrated that a convolutional network can achieve better performance over a recurrent network while being more than 10 times smaller. The local perception and weight sharing of CNN can greatly reduce the number of parameters, thus improving the performance and efficiency of the model.\\

These have motivated extending the easily parallelizable convolutional models for more efficient time series analysis tasks. More importantly, convolutional models can effectively extract multi-scale localized spatial features with little to no need to pre-processed data \cite{abbas2019comprehensive}, \cite{doering2017convolutional}. \cite{oil} proved that convolutional models with 2-dimensional input actually do perform better for short-term prediction than other feed-forward networks neural network models with vector input. To this end, many researchers transform price time-series data into images and then extract multi-scale temporal features using CNN to predict stock trends \cite{candlestick}. Deep convolutional networks show remarkable performance in prediction accuracy for long-range stock trading period \cite{kusuma2019using}, and in some cases, for short-term price forecasting \cite{oil}. Predicting the stock prices in the short-term range relies on short-term trends while getting the long-term range prediction relies on discovering different trading patterns. By careful design of a convolutional network, it is achievable to do both. Despite many published works that adopt convolutional networks for trend prediction, the existing CNN architectures are rather unsophisticated and unrefined.\\

We use intra-correlated market time-series from daily historical stock prices, including open, high, low, and close prices for a timeframe of T trading days. These short-term price frames are then fed to parallel deep convolutional encoders to extract local temporal features in multiple scales. In order to capture the long-term dependency, we fuse same-scale features from consecutive price frames before gradually upsampling these deep coarse feature maps back to the full-size segmentation map and consequently make predictions on the up/down trends. This way, the short-term price frame patterns are extracted from the input frames, and at the same time, long-term dependency is accounted for by incorporating multiple price frames.\\

We compared our proposed architecture with other deep learning models for stock market change trend forecasting based on historical prices and showed the superb performance of our network over previous recursive and feed-forward models. We also experimented with several other fully convolutional networks for semantic segmentation to show the efficiency of both our model and other similar 2D-FCN models for this task. Lastly, with a small modification, we achieved different interesting patterns in segmentation maps for the next 20 days of trend forecasting. The main contributions of this paper can be summarized as follows.\\

\begin{enumerate}
    \item		This paper first proposes a two-dimensional intra-daily price matrix to simulate the image input style to perform semantic segmentation tasks. The final segmentation mask corresponds to the up/down trend of all the next T days' open, high, low, and close prices.\\
    
    \item 	We propose a fully convolutional encoder-decoder network that extracts and incorporates short-term and long-term patterns effectively. We use ASPP blocks to better extract the patterns with different receptive fields, and multiple levels of down-sampling to gain deep coarse feature stacks from multiple consecutive price frames.  After fusing extracted features from parallel encoder streams, a symmetrical up-sampling has been performed to reach the full-scale segmentation map for the next T days prices' change directions. \\

    \item We conducted a comprehensive experiment to study the output dense classification patterns by incorporating additional input frames. We found that not only do the overall results improve with an increased input time horizon, but the pattern also changes, resulting in the most accurate day being pushed further toward the middle of the output timeframe.\\
    
    \item  	To demonstrate the performance of the proposed network, the experimental results are evaluated from two aspects: trend prediction evaluation and semantic segmentation evaluation. In trend prediction performance evaluation, we compare the proposed model with several deep learning-based trend prediction methods. For the latter, we experiment with other infamous fully convolutional semantic segmentation networks to jointly predict future daily prices. In both aspects, our proposed architecture highly outperforms other approaches.\\
\end{enumerate}

\section{Related Works}
\subsection{Deep Learning in Financial Time-series Trend Prediction}
The scientific community generally utilizes two ways for stock market prediction \cite{puneeth2021comparative}. The first approach is the fundamental analysis, where underlying internal and external factors that affect the value of a stock or a company are used as predictive attributes. These factors include the company’s financial performance, social and political behavior, and economic data \cite{beyaz}. The second one is the technical analysis, where the predictive attributes are mainly historical prices and volumes. This method focuses on an analysis of trends in securities’ prices such as daily opening, high, low, and closing prices. Technical analysis is the most common approach in the literature \cite{nazario2017literature}. Since all the new information, like news and macroeconomic variables, are already represented in stock prices, technical analysts believe market price movements tell everything. Therefore, their strategies are based on the stock prices and technical indicators such as relative strength index (RSI) and moving average \cite{nti2020systematic}.\\

Many machine learning \cite{knn, svm1, svm2, ml1} and deep learning \cite{hu2021survey, jiang2021applications, thakkar2021comprehensive} approaches have been proposed to analyze financial time-series for market prediction. The machine learning approaches usually have limited interpretability, need manual feature selection, and perform weakly for very complex tasks. This encourages the integration of deep learning-based models to enhance stock market predictions. The complex intrinsic patterns of stock price trends can be studied using such models by extracting essential characteristics of highly unstructured financial data. Among the deep learning models, Deep ANN and MLP, RNN, LSTM, and CNN have been the dominant models \cite{sezer2020financial}.\\

Deep Artificial Neural Networks and Multi-Layer Perceptron (MLP) have been shown to have superior performance over traditional models \cite{Prime2020ForecastingTC}. \cite{144} used a deep ANN and open, close, high, and low daily prices of the last 10 days of index data. In addition, MLP and ANN were used for the prediction of index data. \cite{142} created an ensemble network of several deep ANN models for trend prediction.  To better handle temporal data, RNN and LSTM are provided as an enhancement of feed-forward neural networks. \cite{jarrah2019recurrent} applied discrete wavelet transform on stock price time-series followed by a deep RNN to predict the closing price of the next 7 days. \cite{186} compared 3 different recurrent models, namely, vanilla RNN, LSTM, and GRU, to predict the movement of stock prices. \cite{133} used LSTM to predict the trend of stock prices and compared the direction of change classification performance with classical time-series forecasting techniques.\\


\subsection{Deep Fully Convolutional Networks}

The idea of dismantling fully connected layers from convolutional layers initially studied in \cite{fcn}, proposing the FCN. The primary objective was to create semantic segmentation map by adapting image classification networks such as AlexNet \cite{alexnet} and VGG \cite{vgg} into fully convolutional networks. The resulting network is considered revolutionary in several aspects. Due to the fully convolutional nature, inference was seen to be considerably faster. More importantly, segmentation maps are allowed to be generated for images with any resolution. And most importantly, they proposed the skip architecture for deep convolutional networks. Deep convolutional networks commonly create feature hierarchies by down-sampling feature maps. Skip connections are used preceding the down-sampling layers to preserve and forward this information to deeper layers allowing information to flow, which would otherwise be lost. This idea to use skip connections in deep convolutional networks eventually evolved into the encoder-decoder structures \cite{segnet, unet} for semantic segmentation. Numerous fully convolutional networks are proposed in this domain, among which we are interested in two types of them that are designed to better capture multi-scale context. \\

A successful encoder-decoder network application is often found in computer vision tasks, such as human pose estimation and object detection. The encoder-decoder network usually consists of two parts: an encoder that gradually reduces the feature maps, while also capturing semantic information, and a decoder that gradually recovers the object details and spatial information. DeconvNet \cite{deconv}, Seg-Net \cite{segnet}, and U-Net \cite{unet} are very well-known examples. They comprise two parts, a contracting path to capture context, and a symmetric expanding path that enables precise localization. However, the highly correlated semantic information, which is provided by the adjacent lower resolution feature map of the encoder, must pass through multiple intermediate layers in order to reach the same decoder stage. This often results in a level of information decay. U-Net overcame this short-coming by utilizing a symmetric skip architecture to provide links between the same-level encoder and decoder stages. Since then, U-Net has been the backbone of many networks for the segmentation of different applications, such as medical images, street view images, satellite images, just to name a few. \\

SegNet utilizes only pooling indices in each stage of the encoder to perform nonlinear up-sampling of the corresponding decoder stage. An issue with encoder-decoder based models is that some of the finer details of an image may be lost due to being part of the resolution that gets lost during encoding. HRNet \cite{hrnet} avoids losing such exact information by connecting the high-to-low resolution convolution streams in a parallel manner and frequently exchanging information between resolutions. They incorporated the multi-scale feature maps in the output node to prevent losing objects at different scales. In fact, many multi-scale models have been proposed. \\

Except using multi-scale features in different stages, many models proposed to also consider multi-scale feature extraction at each stage. Feature Pyramid Network \cite{lin2017feature} is a well-known multiscale analysis model which has been used in different neural network architectures, primarily for object detection but it has also been applied to segmentation. Essentially, the pyramidal hierarchy of deep CNNs is harnessed to create feature pyramids with minimal additional cost. At the same time, to integrate both high- and low- resolution features, the FPN consists of bottom-up and top-down pathways with lateral connections. The spatial pyramid pooling layer adopts parallel convolutional layers with different kernel sizes, then joins the pooled feature maps to fuse feature maps at multiple scales. SPP can effectively increase the extraction range of the backbone features, signiﬁcantly separate the essential contextual features. PSPNet \cite{zhao2017pyramid} applies SPP at several grid scales including image-level pooling. Even though rich semantic information is encoded in the last feature map, detailed information related to object boundaries is missing due to the pooling or convolutions with striding operations within the network backbone. This could be alleviated by applying the atrous convolution to extract denser feature maps. \\

Deeplab V2 \cite{deeplabv2} proposed Atrous Spatial Pyramid Pooling as the feature extraction modules followed by fully connected CRFs. ASPP combines atrous convolutions with different dilation rates in parallel, which can provide multiscale denser contextual information, a larger receptive field, and more local features. Subsequently, \cite{deeplabv3} proposed to omit the CRFs and proposed a deep convolutional network with cascaded and parallel modules of atrous convolutions. And eventually, DeepLabV3+ \cite{deeplabv3+} upgraded the previous design with an encoder-decoder architecture in conjunction with ASPP blocks or Xception \cite{xception} modules. \\

\section{Dataset}

\paragraph{Rescaling} In order to build the input price matrixes, a rescaling of the real observations of the time series stock price is needed so that it falls into a small specific interval. Let $X = {x_1,x_2,...,x_n}$ be the considered time series with $n$ components, the rescaling to the interval $[0, 1]$ is achieved by scaling the maximum value of each time series to unit size.
\begin{equation}
    {\tilde{x_i}} =\frac{x_i - min(X)}{max(X) - min(X)}
\end{equation}
Hence, the scaled series is represented by $\tilde{X} = \{\tilde{x_1}, \tilde{x_2}, ..., \tilde{x_n}\}$

\paragraph{Historical Price Image} For a specific stock, the input frames should contain the lowest, opening, highest, and closing prices for T consecutive trading days. Each row has the rescaled historical prices for a day. The following demonstrates the input matrix, or, price frames.\\

\includegraphics[scale=0.22]{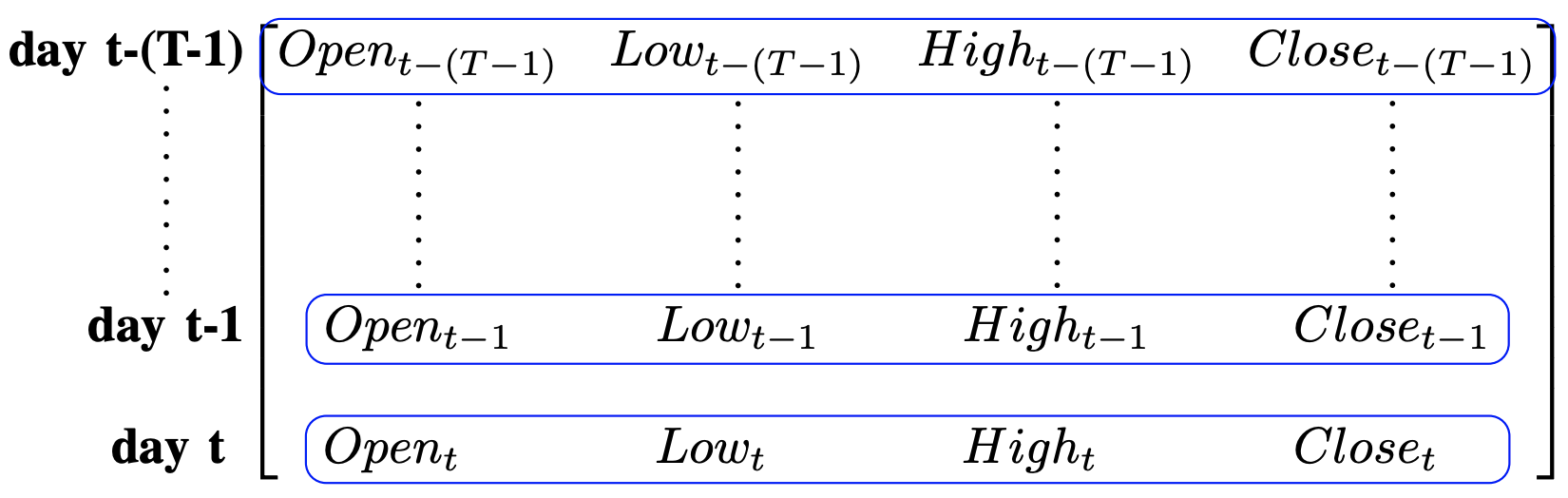} 

\paragraph{Segmentation Mask} To create the annotated labels, we compare the historical price frame at time frame $n+1$ with the one at time frame $n$. More specifically, each component of the stock price matrix for the next time frame is compared with the one from the last time frame, and each pixel is assigned with zero or one according to the following criteria:

$$
{y^{n+1}_{t,c}} = 
    \begin{cases}
    1, & \qquad {X^{n+T}_{t,c}} > {X^n_{t,c}} \\
    0, & \qquad o.w.
    
    \end{cases}
$$

\begin{figure}[H]
    \centering
    \includegraphics[width=\linewidth]{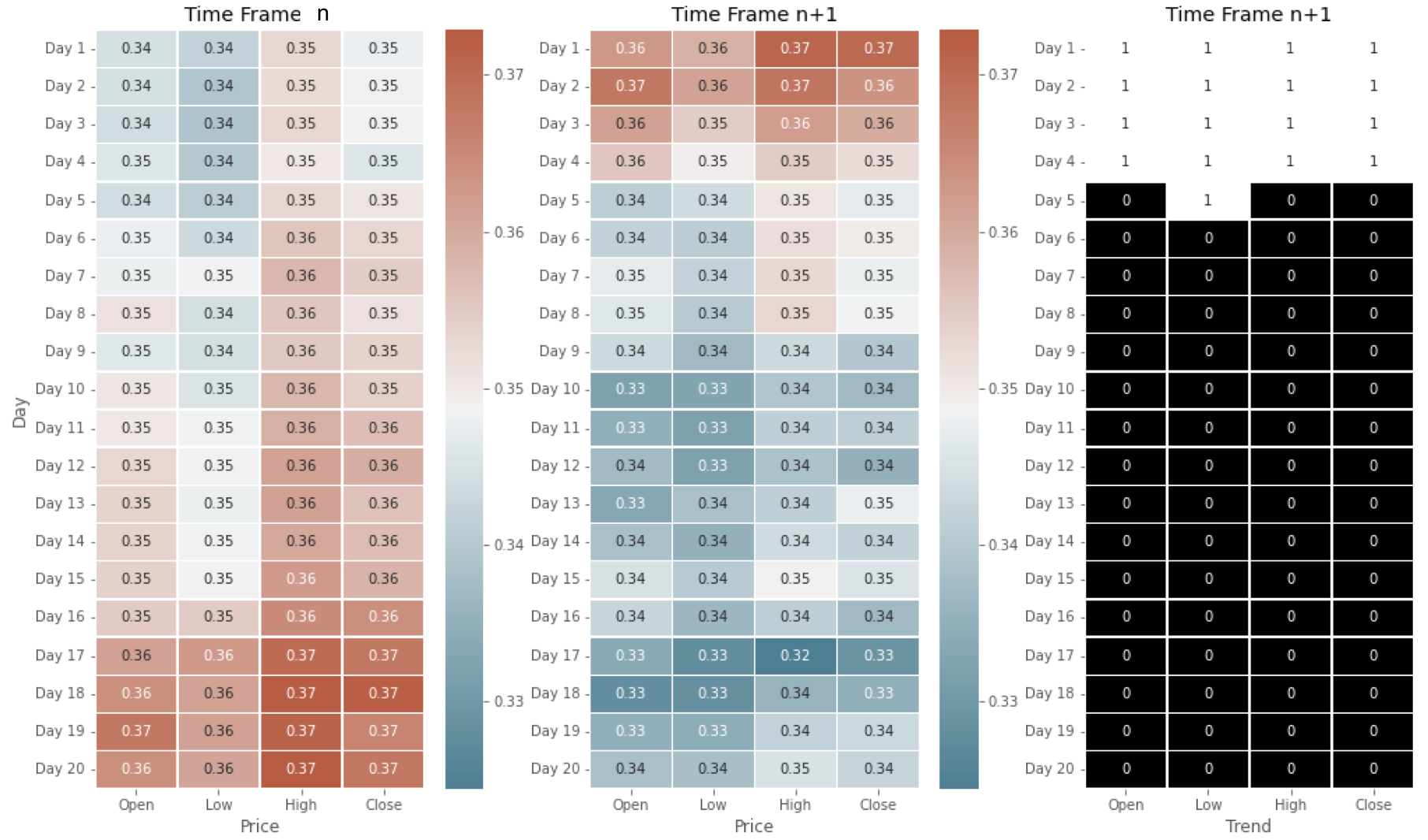}
     \caption{Stock price frames and corresponding output trend segmentation map}
     \label{datamat}
\end{figure}

Where ${y^{n+1}_{t,c}}$ is the stock price trend for the next time frame at row $t$ and column $c$. Resulting matrices are illustrated in Fig. \ref{datamat}.\

\section{Methodology}
We propose a 2D convolutional encoder for extracting hierarchical latent representations from raw daily stock prices. Our approach is designed to capture both short-term patterns and long-term changes in the data, which are essential for accurate prediction. To accomplish this, we make use of multiple time frames inputs and parallel encoders. The parallel encoders allow us to properly encode each time frame based on its specific number. Having different encoder, or parallel encoders is the popular choice for video next-frame prediction models. Our encoder employs Atrous convolutions blocks in multiple stages to downsample the input frames, and the decoder gradually upsamples the concatenated same-stage decoded data for each frame. The decoder is essential for preserving the diligently encoded data and for predicting a fine segmentation map. It is designed to be highly effective, ensuring that the encoded information is not lost and that the final segmentation map is accurate.

\begin{figure}
    \centering
    \includegraphics[width=\linewidth]{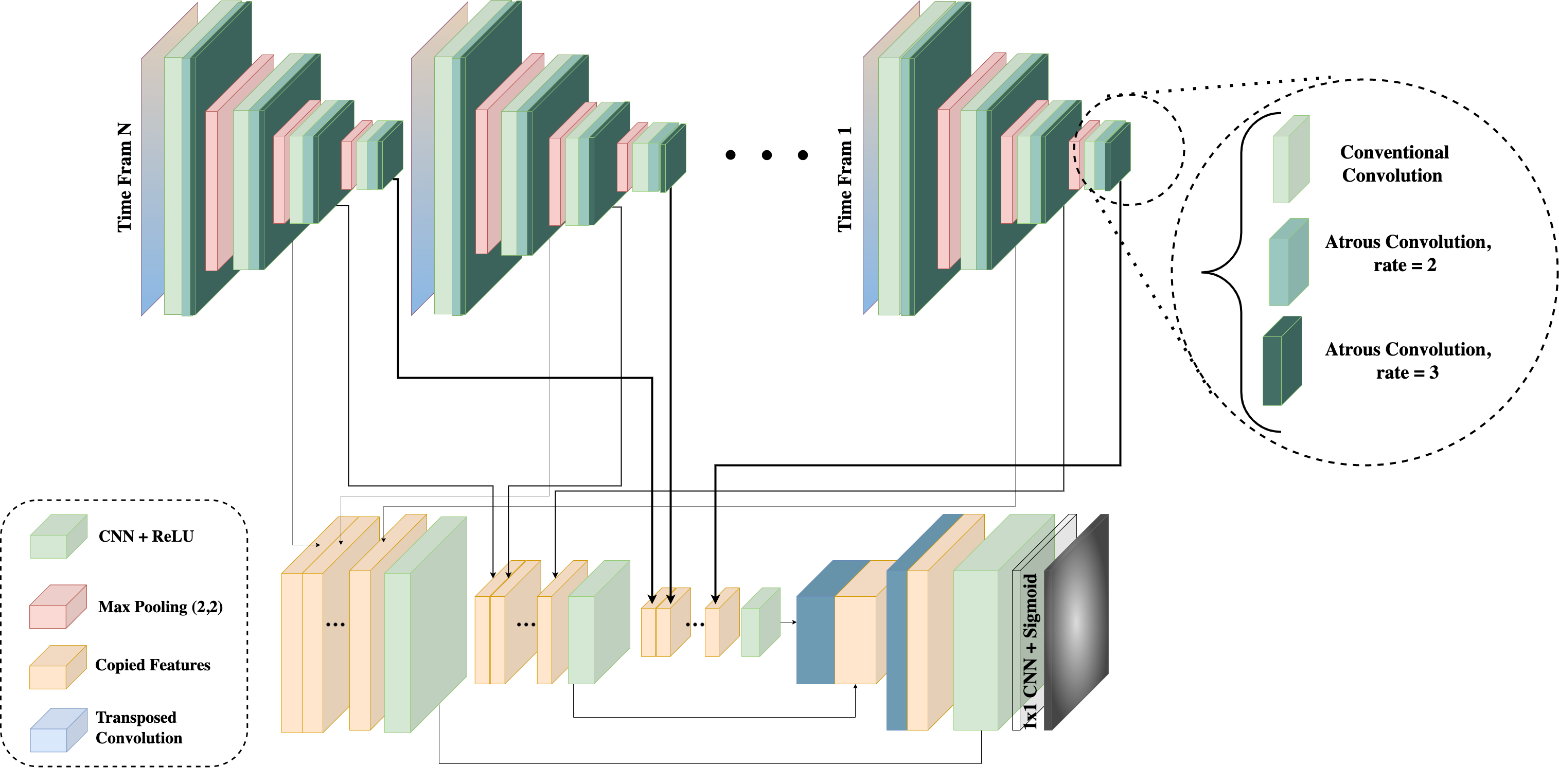}
     \caption{Propose framework. Green stacks represent the ASPP blocks. The N streams of multi-scale features are concatenated to form three scales of feature stacks }
     \label{net}
\end{figure}

\subsection{Atrous Spatial Pyramid Pooling}

\begin{figure}
\includegraphics[width=.45\textwidth]{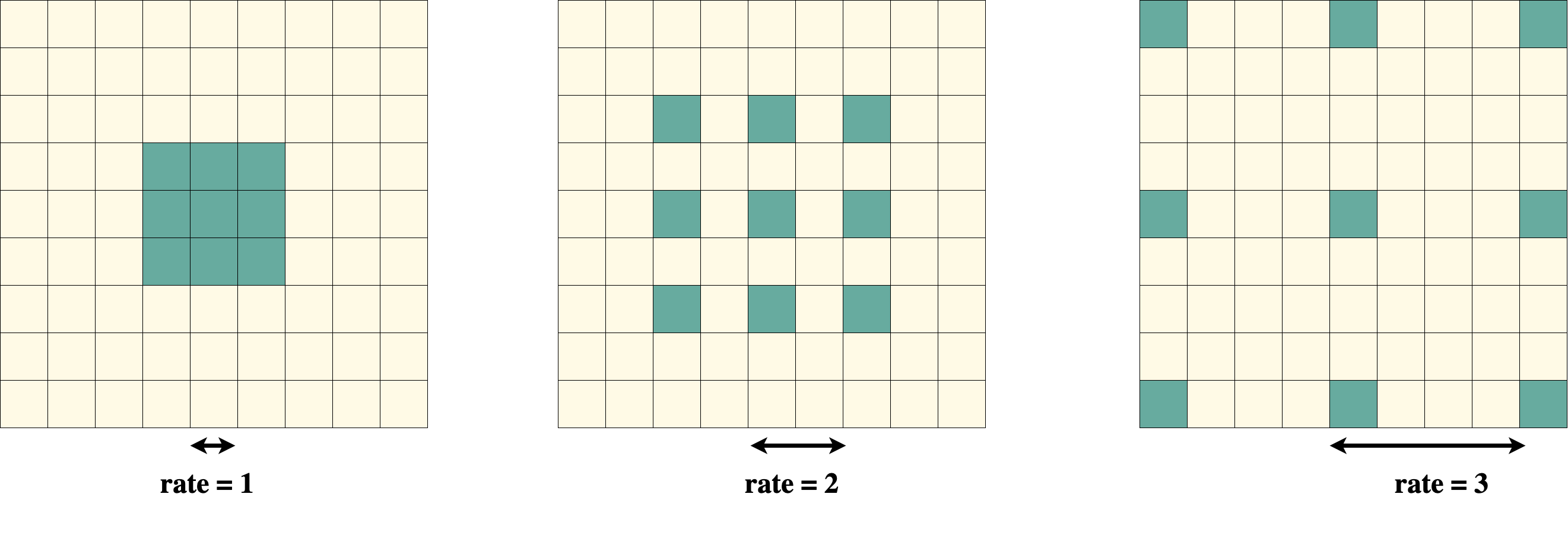}\hfill
\caption{Atrous convolutions with 3x3 kernel and 9x9 feature map}
\label{atrous}
\end{figure}

Atrous convolution is a generalization of standard convolution which allows for explicit control of feature resolution and filter field-of-view in deep convolutional neural networks, enabling the capture of multi-scale information. For two-dimensional feature map $x$ and a filter $f$, each location $i$ on the output $y$ can be defined as:
\begin{equation}
y[i] = \sum_{j} x[i + r \times j] f[j] 
\end{equation}

where $r$ is the dilation or atrous rate. In another word, the dilation rate determines the sampling rate of the input signal, which is similar to convolving the input x with filters that have been upsampled by adding $r - 1$ zeros between each filter value in each dimension. Regular convolution is equivalent to atrous convolution when the dilation rate is set to 1, and adjusting the filter's field of view is achieved by increasing the rate value. Fig. \ref{atrous} illustrates atrous convolutions with different dilation rates.

We employ Atrous Spatial Pyramid Pooling as encoder building blocks. The ASPP, which was introduced in \cite{deeplab}, employs the use of four parallel atrous convolutions with different rates on the feature map. This method builds upon the success of spatial pyramid pooling \cite{spp} by resampling features at various scales to classify regions of arbitrary scale accurately and efficiently. ASPP with its various atrous rates effectively captures multi-scale information, however, as the sampling rate increases the number of valid filter weights decreases \cite{deeplabv3}. Hence, we set the maximum rate to 3, creating a pyramid of 3 feature stacks with dilation rates of 1,2, and 3.

\subsection{Parallel Multi-Scale Dense Feature Extraction}
Let $F = \{f_1, f_2, …, f_N\}$ denote a sequence of $N$ consecutive frames, $f_i \in \mathbb{R}^{T \times 4}$, containing daily prices for T days each. We learn hierarchical latent representation for $f_i$s separately and in parallel. Due to the complicated nature of stock market data, boosting feature extraction power is imperative. Thus, in each stage, we employ ASPP blocks which we will investigate shortly. After each ASPP block the height and maps of the feature stacks are reduced to half their size, resulting in three scales of the feature map. Fig. \ref{net} illustrates the network architecture.\

Conceptually, filters from the lower layers capture primitive local context information while higher-level blocks capture more complicated relationships and coarse features. Such a bottom-up architecture models multi-variate time series data by hierarchical embedding in coarse 3D feature stacks. After encoding these $f_i$s in parallel, we apply the convolution operation to the concatenated feature maps formed by concatenating same-scale features stacks from all $f_i$s, resulting in feature maps $h_1, h_2, h_3$.

\subsection{Transposed Convolutions and Skip Architecture}

Inspired from \cite{unet}, we concatenate the same-scale encoded features with upsampled feature stacks in each decoder stage. This approach helps to prevent loss of information by combining the deep coarse features with the same-scale features from the encoder. In more detail, the deep coarse feature stacks are upsampled using transposed convolutions with strides of 2. The resulting feature stack is then concatenated with the same-scale feature stack from the encoder, followed by another transposed convolution with a stride of 2. This process is repeated again to produce the full-resolution feature maps, and finally, a 1 x 1 standard convolution with sigmoid activation is applied to acquire the final segmentation mask. This approach has been proven to be highly effective in preserving important features and improving the overall performance of the model.


\section{Results}
In this section, we comprehensively evaluate the performance of multiple semantic segmentation models and our proposed framework for predicting stock trends. Furthermore, we analyze the influence of the size of the input frames and the number of frames used on the accuracy of the predictions for various output time horizons. We utilized daily price time-series for five companies from 1970-01-02 to 2017-11-10, resulting in 600 and 300 frames for T of 20 and 40, respectively.

Each stock was trained and tested separately, with a dataset division of 65\%, 10\%, and 25\% for training, validation, and testing. All models were implemented in Keras and trained on T4 GPU with 16GB of memory, utilizing binary cross-entropy as the loss function and Adam optimizer.

\subsection{Semantic Segmentation Models}
\begin{figure}
    \centering
    \subfloat[\centering 20 days frame input and 20 days frame output]{{\includegraphics[scale=0.45]{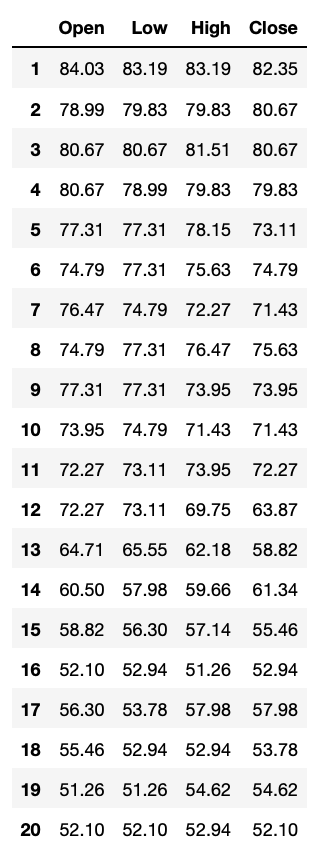} }}%
    \qquad
    \subfloat[\centering 40 days frame input and 20 days frame output]{{\includegraphics[scale=0.45]{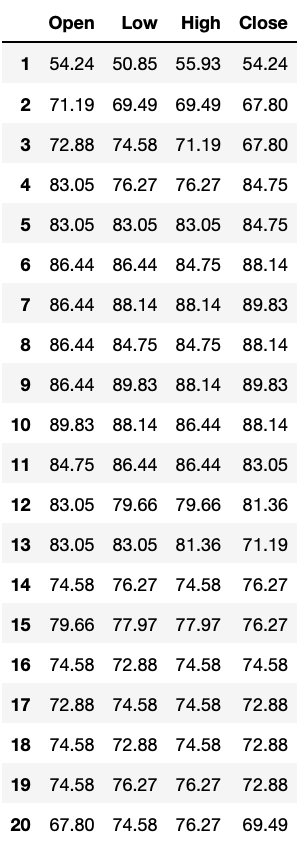} }}%
    \caption{Accuracy maps for networks with different input frame sizes.}%
    \label{maps}%
\end{figure}

In this study, we evaluated the performance of our semantic segmentation model by changing the input-to-output ratio. Initially, we started with an equal input and output frame size, which is an intuitive choice for semantic segmentation. We observed the changing pattern in performance over the whole output horizon and 4 prices. To further investigate this, we visualized the average accuracy of the segmentation maps per pixel. Fig. \ref{maps} shows the accuracy for each price and day, where network (a) takes a 20 x 4 frame as input, and network (b) takes a 40 x 4 frame as input but outputs a 20 x 4 prediction for the next 20 days.

In (a), Although the results were promising for the first few days, with an accuracy of 84.03\% for the first day, it gradually declined to about 52 \% for the final days. However, when we increased the input-to-output ratio to 2, the accuracy improved significantly. On the first day, the accuracy was 54\%, but it quickly increased to 69\% on the second day, reaching its peak at 88\% on the 10th day, before declining to 72\% on the last day. This increase in accuracy can be attributed to the network's ability to capture the "zoomed-out" perspective of the 20-day change patterns with the increased length of input frames. Both models showed impressive total accuracy, but the model with an input-to-output ratio of 2 exhibited the most consistent accuracy across all days and higher maximum accuracy.

Additionally, we compared our results to other semantic segmentation networks in the field. To the best of our knowledge, the application of these models for financial time-series classification has not been explored previously. In an attempt to enhance the network's ability to capture patterns in each frame, we experimented with inputting trends instead of prices. The input trends frames were created similar to the labels and were used instead of the prices frames. We compared the performance of 6 different models, including FCN, SegNet, U-Net, DeepLab V3+, and two proposed models with 20 and 40 days input frames respectively. The results of this comparison are shown in Table \ref{sem}. The table summarizes the AUC and Accuracy of each model when using both price and trend frames as input. 

When analyzing the results for price frames, the proposed-40 model stands out with the highest AUC of 0.88 and accuracy of 78.18\%. The U-Net and FCN models follow closely with AUCs of 0.80 and 0.79 and accuracy scores of 70.77\% and 70.25\%, respectively. The proposed-20 model and DeepLab V3+ also performed well, achieving AUCs of 0.78 and 0.77 and accuracy scores of 68.21\% and 70.32\%, respectively. On the other hand, SegNet had the weakest performance with an AUC of 0.58 and an accuracy of 51.56\%.
For trend input, the FCN model demonstrated the best results with an AUC of 0.77 and an accuracy of 70.77\%. The U-Net model was close behind with an AUC of 0.76 and an accuracy of 69.29\%. The proposed-20 and proposed-40 models still performed relatively well, with AUCs of 0.72 and 0.71 and accuracy scores of 62.82\% and 65.28\%, respectively. The decrease in performance when switching to trend input was minor for FCN and U-Net, likely due to their ability to incorporate similar features from the early encoding stages into the final layers, making learning more effective when input and output are similar.

\begin{table}
    \centering
    \caption{Comparison with other semantic segmentation encoder-decoder architectures}
    \begin{tabular}{l*{4}{c}}
        \toprule
        \multirow{2}{*}{} & \multicolumn{2}{c}  {Prices} &  \multicolumn{2}{c}  {Trends}\\

          Model         &  AUC &  Accuracy    & AUC &  Accuracy\\
        \midrule
         FCN & 0.79 & 70.25\% & 0.77 & 70.77\%  \\ 
         SegNet &      0.58 & 51.56\% & 0.52 & 52.23\% \\ 
         U-Net &       0.80 & 70.77\% & 0.76 & 69.29\% \\
         DeepLab V3+ & 0.77 & 70.32\% & 0.67 & 57.16\% \\ 
         Proposed-20 & 0.78 & 68.21\% & 0.72 & 62.82\% \\ 
         Proposed-40 & 0.88 & 78.18\% & 0.71 & 65.28\% \\ 
         
        \bottomrule  
    \end{tabular}
\label{sem}
\end{table}

\subsection{Input Time Horizon}
We experimented with changing the number of input frames to see the changes in evaluation metrics when the input time horizon increases. We systematically varied the number of input frames from 1 to 11, and used 40-days input frames. Our findings suggest that with only 1 input frame, the network's behavior follows the typical pattern observed in CNN models, exhibiting better results for short-term predictions. As the number of input frames increases, the longer-term results improve, a phenomenon reminiscent of the behavior of recursive networks.

The results reveal that the best overall model is the one with 9 input frames (corresponding to 360 days), achieving an AUC of 0.88 and an accuracy of 88\% in total. For first-day predictions, models with 1-2 input frames perform the best, with a rapid decline in accuracy and AUC as the number of input frames increases. For day 5, models with 2 frames input perform the best, with the 9-frame model also exhibiting an impressive AUC of 90\% and the fastest growth among long-term input models. For day 10, the 9-frame model reaches the highest AUC and accuracy among all models and days in the table, with an AUC of 0.95, an accuracy of 88\%, and an F1 score of 0.89. For day 20, the best overall metrics are achieved by the 7-frame input model, with an accuracy of 77\% and an F1 score of 0.78. The 9-frame input model, however, results in a higher AUC of 0.85.

In conclusion, while among the models with a higher number of input frames, the 9-frame model performs the best overall, among the models with a lower number of input frames, the 2-frame model performs the best, although the single-frame model has close results. Eventually, this shows that it’s possible to predict the next few days' trends with very high accuracy using semantic segmentation with only the last 1-2 trading months input.  It also confirms that short-term relationships are decent predictors for the next 1-2 days' trends. With an increasing number of input frames, the maximum accuracy increases in value and moves to further days. Except for 1 and 2 frames models, for day 1, the other models do not perform well. This probably is due to incorporating long-term information for the 1-2 days, which is not much of use for those cases. However, as can be seen in Figure \ref{maps} (b) it improves really fast on days 2-3 for the 9-frame version.


\begin{table*}[] \centering
\begin{small}
\begin{tabular}{@{}lrrrrrrrrrrrr@{}}\toprule
No. of Frames & \textbf{1} & \textbf{2} & \textbf{3} & \textbf{4} & \textbf{5} & \textbf{6} & \textbf{7} & \textbf{8} & \textbf{9}& \textbf{10}  & \textbf{11}\\ \midrule

\textbf{Day 1} &  &  & &  &  &  &  &  & \\ \midrule
\textbf{AUC}	 & 	\textbf{0.90}	 & 	0.88	 & 	0.66	 & 	0.51	 & 	0.59	 & 	0.46	 & 	0.57	 & 	0.39	 & 	0.54	 & 	0.49	 & 	0.57\\    \hdashline
\textbf{Acc. (\%)}	 & 	74.17	 & 	\textbf{75.02}	 & 	56.25	 & 	52.50	 & 	53.75	 & 	49.15	 & 	48.73	 & 	45.76	 & 	53.81	 & 	48.73	 & 	60.34\\    \hdashline
\textbf{Precision}	 & 	0.68	 & 	\textbf{0.69}	 & 	0.53	 & 	0.51	 & 	0.52	 & 	0.50	 & 	0.49	 & 	0.47	 & 	0.53	 & 	0.49	 & 	0.57\\    \hdashline
\textbf{Recall}	 & 	0.92	 & 	0.91	 & 	\textbf{0.98}	 & 	0.86	 & 	0.83	 & 	0.50	 & 	0.70	 & 	0.61	 & 	0.86	 & 	0.73	 & 	0.83\\    \hdashline
\textbf{f1}    &		\textbf{0.78}    &	\textbf{0.78}    &	0.69    &	0.64    &	0.64    &	0.50    &	0.58    &	0.53    &	0.65    &	0.59    &	0.68\\

\midrule

\textbf{Day 5} &  &  & &  &  &  &  &  & \\ \midrule
\textbf{AUC}	 & 	0.90	 & 	\textbf{0.92}	 & 	0.73	 & 	0.75	 & 	0.73	 & 	0.80	 & 	0.88	 & 	0.85	 & 	\textbf{0.92}	 & 	0.83	 & 	0.89\\    \hdashline
\textbf{Acc. (\%)}	 & 	82.08	 & 	\textbf{87.50}	 & 	62.50	 & 	65.00	 & 	72.50	 & 	76.69	 & 	84.75	 & 	79.66	 & 	83.47	 & 	72.46	 & 	81.03\\    \hdashline
\textbf{Precision}	 & 	0.83	 & 	\textbf{0.89}	 & 	0.60	 & 	0.63	 & 	0.71	 & 	0.83	 & 	0.82	 & 	0.85	 & 	0.86	 & 	0.72	 & 	0.78\\    \hdashline
\textbf{Recall}	 & 	0.83	 & 	0.87	 & 	0.89	 & 	0.81	 & 	0.83	 & 	0.69	 & 	\textbf{0.90}	 & 	0.74	 & 	0.81	 & 	0.77	 & 	\textbf{0.90}\\    \hdashline

\textbf{f1}    &	0.83    &	\textbf{0.88}    &	0.72    &	0.71    &	0.76    &	0.76    &	0.86    &	0.79    &	0.84    &	0.75    &	0.84 \\
         
\midrule

\textbf{Day 10} &  &  & &  &  &  &  &  & \\ \midrule
\textbf{AUC}	 & 	0.89	 & 	0.84	 & 	0.87	 & 	0.91	 & 	0.90	 & 	0.94	 & 	0.93	 & 	0.94	 & 	\textbf{0.95}	 & 	0.94	 & 	0.92\\    \hdashline
\textbf{Acc. (\%)}	 & 	82.50	 & 	73.33	 & 	77.08	 & 	85.42	 & 	87.08	 & 	80.08	 & 	85.17	 & 	86.86	 & 	\textbf{88.14}	 & 	87.71	 & 	83.19\\    \hdashline
\textbf{Precision}	 & 	0.81	 & 	0.79	 & 	0.73	 & 	0.86	 & 	\textbf{0.89}	 & 	0.79	 & 	0.86	 & 	\textbf{0.89}	 & 	\textbf{0.89}	 & 	0.88	 & 	0.82\\    \hdashline
\textbf{Recall}	 & 	0.87	 & 	0.69	 & 	\textbf{0.92}	 & 	0.87	 & 	0.87	 & 	0.87	 & 	0.87	 & 	0.87	 & 	0.89	 & 	0.90	 & 	0.89\\   \hdashline
\textbf{f1}	    &	0.84	    &	0.74    &		0.81    &		0.87    &		0.88    &		0.82    &		0.86    &		0.88    &		\textbf{0.89}    &		\textbf{0.89}    &		0.85 \\
\midrule

\textbf{Day 20} &  &  & &  &  &  &  &  & \\ \midrule
\textbf{AUC}	 & 	0.71	 & 	0.77	 & 	0.81	 & 	0.80	 & 	0.83	 & 	0.84	 & 	0.84	 & 	0.84	 & 	\textbf{0.85}	 & 	0.83	 & 	0.80\\    \hdashline
\textbf{Acc. (\%)}	 & 	57.08	 &	65.41	 &	64.17	 &	65.83	 &	75.00	 &	75.42	 &	\textbf{77.54}	 &	74.58	 &	72.03	 &	73.73	 &	69.83\\    \hdashline
\textbf{Precision}	 & 	0.54	 & 	0.61	 & 	0.59	 & 	0.63	 & 	0.75	 & 	0.71	 & 	\textbf{0.76}	 & 	0.73	 & 	0.68	 & 	0.69	 & 	0.66\\    \hdashline
\textbf{Recall}	 & 	\textbf{0.95}	 & 	0.89	 & 	0.93	 & 	0.82	 & 	0.76	 & 	0.85	 & 	0.81	 & 	0.77	 & 	0.85	 & 	0.86	 & 	0.83\\ \hdashline
\textbf{f1}	    & 	0.69	& 0.72	& 0.73   & 	0.71   &	0.76	 &  \textbf{0.78}	&  \textbf{0.78}  &	0.75 & 	0.75 &	0.77  &	0.74\\

\toprule

\textbf{Total} &  &  & &  &  &  &  &  & \\ \midrule
\textbf{AUC}	 & 	0.85	 & 	0.84	 & 	0.80	 & 	0.81	 & 	0.81	 & 	0.82	 & 	0.85	 & 	0.85	 & 	\textbf{0.88}	 & 	0.86	 & 	0.84\\    \hdashline
\textbf{Acc. (\%)}	 & 	73.87	 & 	74.90	 & 	67.98	 & 	72.65	 & 	74.63	 & 	74.85	 & 	77.61	 & 	77.06	 & 	\textbf{78.18}	 & 	76.02	 & 	75.13\\    \hdashline
\textbf{Precision}	 & 	0.69	 & 	0.74	 & 	0.64	 & 	0.70	 & 	0.74	 & 	0.75	 & 	0.76	 & 	\textbf{0.78}	 & 	0.76	 & 	0.74	 & 	0.72\\    \hdashline
\textbf{Recall}	 & 	0.88	 & 	0.81	 & 	0.90	 & 	0.82	 & 	0.80	 & 	0.77	 & 	0.82	 & 	0.76	 & 	\textbf{0.85}	 & 	0.82	 & 	0.84\\    \hdashline

\textbf{f1}	    &	0.77    &	0.77    &	0.75    &	0.76    &	0.77    &	0.76    &	0.79    &	0.77    &	\textbf{0.80}    &	0.78    &	0.78 \\

\bottomrule
\end{tabular}
\end{small}
\caption{ Short-term to long-term. The effect of increasing the number of input frames on different days in the output frame. Days 1, 5, 10, and 20 are picked for comparison. }
\end{table*}

\subsection{Deep Learning Stock Trend Prediction Models}
We compared our model with several models designed for stock trend prediction. We chose two models that use CNN as an encoder and then use fully connected and LSTM as decoder units to compare our model with, in conjunction with a Multi-Layer Perceptron model. We used 40-day input (as in Fig. \ref{net}) with 1-frame and 9-frame inputs and changed the prediction horizons of the baseline models. For next-day prediction, among the baselines, Conv-LSTM performs the best, with 63.41\% accuracy. Our proposed 1-frame network predictably performs the best with 75\% accuracy. Our 9-frame network performs similarly to other baselines, with only about 1\% increase in accuracy. 

For the next 4 days' prediction, both versions of the proposed networks have similar results with about 85\% and 0.91 in accuracy and AUC.  While the performance of the Conv-LSTM model decreased to 61.73\%, the other two, interestingly the ones with fully connected layers increased their accuracy to 68.75\% and 55.46\%, for CNN-based one and DNN one respectively. For the next 20 days prediction, the 9-frame proposed network works the best, with 69.69\% and 0.83 in accuracy and AUC, respectively, and CNN+FC is next with 68.75\% and 0.61. Finally, DNN and proposed-1 come next with 60.92\% and 60.33\% in accuracy. As opposed to proposed-1, As the prediction horizon increases, the performance of the DNN model improves.

\begin{table}
\begin{tabular}{cccc}
\toprule
Time Interval  & Model & Acc.(\%) & AUC   \\ 
\midrule
\multicolumn{1}{c}{}           & \multicolumn{1}{c}{CNN+FC \cite{cnn}}    &     51.26    &    0.54       \\
\hdashline

\multicolumn{1}{c}{\relax}             & \multicolumn{1}{c}{CNN+LSTM \cite{convlstm}}     
&    63.41      &     0.49     \\
\hdashline
\multicolumn{1}{c}{1 Day}                               & \multicolumn{1}{c}{DNN \cite{dnn}}   &     52.94     &     0.53     \\ 
\hdashline
\multicolumn{1}{c}{\relax}                               & \multicolumn{1}{c}{Proposed-1}   
&     75.00     &     0.90      \\ 
\hdashline
\multicolumn{1}{c}{\relax}                               & \multicolumn{1}{c}{Proposed-9}   
&     54.24     &     0.54      \\ 
\midrule
\multicolumn{1}{c}{}   & \multicolumn{1}{c}{CNN+FC}     
&     68.75    &   0.64     \\
\hdashline
\multicolumn{1}{c}{}             & \multicolumn{1}{c}{CNN+LSTM}     
&    61.73      &     0.53     \\
\hdashline
\multicolumn{1}{c}{4 Days}                               & \multicolumn{1}{c}{DNN}
&     55.46     &     0.52     \\ 
\hdashline
\multicolumn{1}{c}{\relax}                               & \multicolumn{1}{c}{Proposed-1}   
&     85.01     &     0.92      \\
\hdashline
\multicolumn{1}{c}{\relax}                               & \multicolumn{1}{c}{Proposed-9}   
&     84.75     &     0.90      \\ 
\midrule
\multicolumn{1}{c}{}                & \multicolumn{1}{c}{CNN+FC}     
&     68.75    &    0.61      \\
\hdashline
\multicolumn{1}{c}{\relax}             & \multicolumn{1}{c}{CNN+LSTM}     
&    55.56      &     0.54      \\
\hdashline
\multicolumn{1}{c}{20 Days}                               & \multicolumn{1}{c}{DNN}   &     60.92     &     0.60      \\ 
\hdashline
\multicolumn{1}{c}{\relax}                               & \multicolumn{1}{c}{Proposed-1}   
&     60.33     &     0.74      \\ 
\hdashline
\multicolumn{1}{c}{\relax}                               & \multicolumn{1}{c}{Proposed-9}   
&     69.49     &     0.83     \\ 
\bottomrule
\end{tabular}
\caption{Comparison of the stock trend prediction of different models for different output time horizons}
\label{Fin}
\end{table}

\section{Conclusion}
In this paper, we tackled stock trend prediction in a novel way. We used four daily stock prices to jointly predict their trends for an arbitrary length of time. We introduced the first fully 2D-convolutional model for predicting stock trends based on the semantic segmentation of these price frames. We used Atrous Spatial Pyramid Pooling to extract high-quality features from raw daily prices and encode them in coarse multi-scale feature maps in parallel. Using a deconvolutional decoder, segmentation maps are generated. With increasing parallel encoder modules, or in other words, input time horizon, we show that along with highly accurate overall results, the maximum performance day can move to earlier or later days. We demonstrated that change in encoder input frame size or parallel modules can result in a different accuracy pattern in the segmentation map, one resulting in astonishing short-term prediction accuracy and the other with very impressive long-term predictions. We also compared our model with other deep learning models for different time ranges, proving the superior performance of the proposed network. Experimental results of well-known fully convolutional models for semantic segmentation show that these models have a high potential for the task of stock trend prediction.

\section*{Declarations}
All authors certify that they have no affiliations with or involvement in any organization or entity with any financial interest or non-financial interest in the subject matter or materials discussed in this manuscript.
\subsection*{Funding}
The authors did not receive support from any organization for the submitted work.
\subsection*{Data availability}
The data that supports the findings of this study is publicly available online at https://www.kaggle.com/datasets/borismarjanovic/price-volume-data-for-all-us-stocks-etfs

\bibliography{sn-bibliography}

\begin{thebibliography}{50}
\providecommand{\natexlab}[1]{#1}
\providecommand{\url}[1]{{#1}}
\providecommand{\urlprefix}{URL }
\providecommand{\doi}[1]{\url{https://doi.org/#1}}
\providecommand{\eprint}[2][]{\url{#2}}
 \bibcommenthead

\bibitem[{Abbas et~al(2019)Abbas, Ibrahim, and Jaffar}]{abbas2019comprehensive}
Abbas Q, Ibrahim ME, Jaffar MA (2019) A comprehensive review of recent advances
  on deep vision systems. Artificial Intelligence Review 52(1):39--76

\bibitem[{Anand(2021)}]{anand2021comparison}
Anand C (2021) Comparison of stock price prediction models using pre-trained
  neural networks. Journal of Ubiquitous Computing and Communication
  Technologies (UCCT) 3(02):122--134

\bibitem[{Badrinarayanan et~al(2017)Badrinarayanan, Kendall, and
  Cipolla}]{segnet}
Badrinarayanan V, Kendall A, Cipolla R (2017) Segnet: A deep convolutional
  encoder-decoder architecture for image segmentation. IEEE transactions on
  pattern analysis and machine intelligence 39(12):2481--2495

\bibitem[{Bai et~al(2018)Bai, Kolter, and Koltun}]{bai2018empirical}
Bai S, Kolter JZ, Koltun V (2018) An empirical evaluation of generic
  convolutional and recurrent networks for sequence modeling. arXiv preprint
  arXiv:180301271

\bibitem[{Beyaz et~al(2018)Beyaz, Tekiner, Zeng, and Keane}]{beyaz}
Beyaz E, Tekiner F, Zeng Xj, et~al (2018) Comparing technical and fundamental
  indicators in stock price forecasting. In: 2018 IEEE 20th International
  Conference on High Performance Computing and Communications; IEEE 16th
  International Conference on Smart City; IEEE 4th International Conference on
  Data Science and Systems (HPCC/SmartCity/DSS), IEEE, pp 1607--1613

\bibitem[{Chai et~al(2015)Chai, Du, Lai, and Lee}]{svm1}
Chai J, Du J, Lai KK, et~al (2015) A hybrid least square support vector machine
  model with parameters optimization for stock forecasting. Mathematical
  Problems in Engineering 2015

\bibitem[{Chen et~al(2017{\natexlab{a}})Chen, Papandreou, Kokkinos, Murphy, and
  Yuille}]{deeplab}
Chen LC, Papandreou G, Kokkinos I, et~al (2017{\natexlab{a}}) Deeplab: Semantic
  image segmentation with deep convolutional nets, atrous convolution, and
  fully connected crfs. IEEE transactions on pattern analysis and machine
  intelligence 40(4):834--848

\bibitem[{Chen et~al(2017{\natexlab{b}})Chen, Papandreou, Schroff, and
  Adam}]{deeplabv3}
Chen LC, Papandreou G, Schroff F, et~al (2017{\natexlab{b}}) Rethinking atrous
  convolution for semantic image segmentation. arXiv preprint arXiv:170605587

\bibitem[{Chen et~al(2018)Chen, Zhu, Papandreou, Schroff, and
  Adam}]{deeplabv3+}
Chen LC, Zhu Y, Papandreou G, et~al (2018) Encoder-decoder with atrous
  separable convolution for semantic image segmentation. In: Proceedings of the
  European conference on computer vision (ECCV), pp 801--818

\bibitem[{Chen and Wu(2017)}]{chen2017cnn}
Chen Q, Wu R (2017) Cnn is all you need. arXiv preprint arXiv:171209662

\bibitem[{Cho et~al(2014)Cho, Van~Merri{\"e}nboer, Gulcehre, Bahdanau,
  Bougares, Schwenk, and Bengio}]{cho2014learning}
Cho K, Van~Merri{\"e}nboer B, Gulcehre C, et~al (2014) Learning phrase
  representations using rnn encoder-decoder for statistical machine
  translation. arXiv preprint arXiv:14061078

\bibitem[{Chollet(2017)}]{xception}
Chollet F (2017) Xception: Deep learning with depthwise separable convolutions.
  In: Proceedings of the IEEE conference on computer vision and pattern
  recognition, pp 1251--1258

\bibitem[{Chung et~al(2014)Chung, Gulcehre, Cho, and
  Bengio}]{chung2014empirical}
Chung J, Gulcehre C, Cho K, et~al (2014) Empirical evaluation of gated
  recurrent neural networks on sequence modeling. arXiv preprint arXiv:14123555

\bibitem[{Di~Persio and Honchar(2017)}]{186}
Di~Persio L, Honchar O (2017) Recurrent neural networks approach to the
  financial forecast of google assets. International journal of Mathematics and
  Computers in simulation 11:7--13

\bibitem[{Doering et~al(2017)Doering, Fairbank, and
  Markose}]{doering2017convolutional}
Doering J, Fairbank M, Markose S (2017) Convolutional neural networks applied
  to high-frequency market microstructure forecasting. In: 2017 9th computer
  science and electronic engineering (ceec), IEEE, pp 31--36

\bibitem[{Hansson(2017)}]{133}
Hansson M (2017) On stock return prediction with lstm networks

\bibitem[{He et~al(2015)He, Zhang, Ren, and Sun}]{spp}
He K, Zhang X, Ren S, et~al (2015) Spatial pyramid pooling in deep
  convolutional networks for visual recognition. IEEE transactions on pattern
  analysis and machine intelligence 37(9):1904--1916

\bibitem[{Hu et~al(2018)Hu, Hu, Yang, Yu, Sung, Zhang, Xie, Liu, Robertson,
  Hospedales et~al}]{candlestick}
Hu G, Hu Y, Yang K, et~al (2018) Deep stock representation learning: From
  candlestick charts to investment decisions. In: 2018 IEEE international
  conference on acoustics, speech and signal processing (ICASSP), IEEE, pp
  2706--2710

\bibitem[{Hu et~al(2021)Hu, Zhao, and Khushi}]{hu2021survey}
Hu Z, Zhao Y, Khushi M (2021) A survey of forex and stock price prediction
  using deep learning. Applied System Innovation 4(1):9

\bibitem[{Jarrah and Salim(2019)}]{jarrah2019recurrent}
Jarrah M, Salim N (2019) A recurrent neural network and a discrete wavelet
  transform to predict the saudi stock price trends. International Journal of
  Advanced Computer Science and Applications 10(4)

\bibitem[{Jiang(2021)}]{jiang2021applications}
Jiang W (2021) Applications of deep learning in stock market prediction: recent
  progress. Expert Systems with Applications 184:115,537

\bibitem[{King et~al(2018)King, Bhandarkar, and Hopkinson}]{deeplabv2}
King A, Bhandarkar SM, Hopkinson BM (2018) A comparison of deep learning
  methods for semantic segmentation of coral reef survey images. In:
  Proceedings of the IEEE conference on computer vision and pattern recognition
  workshops, pp 1394--1402

\bibitem[{Krizhevsky et~al(2017)Krizhevsky, Sutskever, and Hinton}]{alexnet}
Krizhevsky A, Sutskever I, Hinton GE (2017) Imagenet classification with deep
  convolutional neural networks. Communications of the ACM 60(6):84--90

\bibitem[{Kusuma et~al(2019)Kusuma, Ho, Kao, Ou, and Hua}]{kusuma2019using}
Kusuma RMI, Ho TT, Kao WC, et~al (2019) Using deep learning neural networks and
  candlestick chart representation to predict stock market. arXiv preprint
  arXiv:190312258

\bibitem[{Labiad et~al(2016)Labiad, Berrado, and Benabbou}]{ml1}
Labiad B, Berrado A, Benabbou L (2016) Machine learning techniques for short
  term stock movements classification for moroccan stock exchange. In: 2016
  11th International Conference on Intelligent Systems: Theories and
  Applications (SITA), IEEE, pp 1--6

\bibitem[{Lin et~al(2017)Lin, Doll{\'a}r, Girshick, He, Hariharan, and
  Belongie}]{lin2017feature}
Lin TY, Doll{\'a}r P, Girshick R, et~al (2017) Feature pyramid networks for
  object detection. In: Proceedings of the IEEE conference on computer vision
  and pattern recognition, pp 2117--2125

\bibitem[{Long et~al(2015)Long, Shelhamer, and Darrell}]{fcn}
Long J, Shelhamer E, Darrell T (2015) Fully convolutional networks for semantic
  segmentation. In: Proceedings of the IEEE conference on computer vision and
  pattern recognition, pp 3431--3440

\bibitem[{Long et~al(2020)Long, Chen, He, Wu, and Ren}]{LONG2020106205}
Long J, Chen Z, He W, et~al (2020) An integrated framework of deep learning and
  knowledge graph for prediction of stock price trend: An application in
  chinese stock exchange market. Applied Soft Computing 91:106,205.
  \doi{https://doi.org/10.1016/j.asoc.2020.106205},
  \urlprefix\url{https://www.sciencedirect.com/science/article/pii/S1568494620301459}

\bibitem[{Luo et~al(2019)Luo, Cai, Tanaka, Takiguchi, Kinkyo, and Hamori}]{oil}
Luo Z, Cai X, Tanaka K, et~al (2019) Can we forecast daily oil futures prices?
  experimental evidence from convolutional neural networks. Journal of risk and
  financial management 12(1):9

\bibitem[{Mehtab and Sen(2020)}]{mehtab2020time}
Mehtab S, Sen J (2020) A time series analysis-based stock price prediction
  using machine learning and deep learning models. arXiv preprint
  arXiv:200411697

\bibitem[{Naz{\'a}rio et~al(2017)Naz{\'a}rio, e~Silva, Sobreiro, and
  Kimura}]{nazario2017literature}
Naz{\'a}rio RTF, e~Silva JL, Sobreiro VA, et~al (2017) A literature review of
  technical analysis on stock markets. The Quarterly Review of Economics and
  Finance 66:115--126

\bibitem[{Noh et~al(2015)Noh, Hong, and Han}]{deconv}
Noh H, Hong S, Han B (2015) Learning deconvolution network for semantic
  segmentation. In: Proceedings of the IEEE international conference on
  computer vision, pp 1520--1528

\bibitem[{Nti et~al(2020)Nti, Adekoya, and Weyori}]{nti2020systematic}
Nti IK, Adekoya AF, Weyori BA (2020) A systematic review of fundamental and
  technical analysis of stock market predictions. Artificial Intelligence
  Review 53(4):3007--3057

\bibitem[{Park and Lee(2016)}]{park2016fully}
Park SR, Lee J (2016) A fully convolutional neural network for speech
  enhancement. arXiv preprint arXiv:160907132

\bibitem[{Prachyachuwong and Vateekul(2021)}]{info12060250}
Prachyachuwong K, Vateekul P (2021) Stock trend prediction using deep learning
  approach on technical indicator and industrial specific information.
  Information 12(6). \doi{10.3390/info12060250},
  \urlprefix\url{https://www.mdpi.com/2078-2489/12/6/250}

\bibitem[{Prime(2020)}]{Prime2020ForecastingTC}
Prime ST (2020) Forecasting the changes in daily stock prices in shanghai stock
  exchange using neural network and ordinary least squares regression.
  Investment Management and Financial Innovations

\bibitem[{Puneeth et~al(2021)Puneeth, Rudagi, Namratha, Patil, and
  Wadi}]{puneeth2021comparative}
Puneeth K, Rudagi S, Namratha M, et~al (2021) Comparative study: Stock
  prediction using fundamental and technical analysis. In: 2021 IEEE
  International Conference on Mobile Networks and Wireless Communications
  (ICMNWC), IEEE, pp 1--4

\bibitem[{Ronneberger et~al(2015)Ronneberger, Fischer, and Brox}]{unet}
Ronneberger O, Fischer P, Brox T (2015) U-net: Convolutional networks for
  biomedical image segmentation. In: International Conference on Medical image
  computing and computer-assisted intervention, Springer, pp 234--241

\bibitem[{Sezer et~al(2020)Sezer, Gudelek, and Ozbayoglu}]{sezer2020financial}
Sezer OB, Gudelek MU, Ozbayoglu AM (2020) Financial time series forecasting
  with deep learning: A systematic literature review: 2005--2019. Applied soft
  computing 90:106,181

\bibitem[{Shynkevich et~al(2014)Shynkevich, McGinnity, Coleman, Li, and
  Belatreche}]{svm2}
Shynkevich Y, McGinnity TM, Coleman S, et~al (2014) Forecasting stock price
  directional movements using technical indicators: investigating window size
  effects on one-step-ahead forecasting. In: 2014 IEEE Conference on
  Computational Intelligence for Financial Engineering \& Economics (CIFEr),
  IEEE, pp 341--348

\bibitem[{Simonyan and Zisserman(2014)}]{vgg}
Simonyan K, Zisserman A (2014) Very deep convolutional networks for large-scale
  image recognition. arXiv preprint arXiv:14091556

\bibitem[{Sinta et~al(2014)Sinta, Wijayanto, and Sartono}]{knn}
Sinta D, Wijayanto H, Sartono B (2014) Ensemble k-nearest neighbors method to
  predict rice price in indonesia. Applied Mathematical Sciences
  8(160):7993--8005

\bibitem[{Song et~al(2019)Song, Lee, and Lee}]{dnn}
Song Y, Lee JW, Lee J (2019) A study on novel filtering and relationship
  between input-features and target-vectors in a deep learning model for stock
  price prediction. Applied Intelligence 49(3):897--911

\bibitem[{Sun et~al(2019)Sun, Xiao, Liu, and Wang}]{hrnet}
Sun K, Xiao B, Liu D, et~al (2019) Deep high-resolution representation learning
  for human pose estimation. In: Proceedings of the IEEE/CVF conference on
  computer vision and pattern recognition, pp 5693--5703

\bibitem[{Thakkar and Chaudhari(2021)}]{thakkar2021comprehensive}
Thakkar A, Chaudhari K (2021) A comprehensive survey on deep neural networks
  for stock market: The need, challenges, and future directions. Expert Systems
  with Applications 177:114,800

\bibitem[{Tsantekidis et~al(2017)Tsantekidis, Passalis, Tefas, Kanniainen,
  Gabbouj, and Iosifidis}]{cnn}
Tsantekidis A, Passalis N, Tefas A, et~al (2017) Forecasting stock prices from
  the limit order book using convolutional neural networks. In: 2017 IEEE 19th
  conference on business informatics (CBI), IEEE, pp 7--12

\bibitem[{Wu et~al(2021)Wu, Li, Herencsar, Vo, and Lin}]{convlstm}
Wu JMT, Li Z, Herencsar N, et~al (2021) A graph-based cnn-lstm stock price
  prediction algorithm with leading indicators. Multimedia Systems pp 1--20

\bibitem[{Yang et~al(2017)Yang, Gong, and Yang}]{142}
Yang B, Gong ZJ, Yang W (2017) Stock market index prediction using deep neural
  network ensemble. In: 2017 36th chinese control conference (ccc), IEEE, pp
  3882--3887

\bibitem[{Yong et~al(2017)Yong, Abdul~Rahim, and Abdullah}]{144}
Yong BX, Abdul~Rahim MR, Abdullah AS (2017) A stock market trading system using
  deep neural network. In: Asian simulation conference, Springer, pp 356--364

\bibitem[{Zhao et~al(2017)Zhao, Shi, Qi, Wang, and Jia}]{zhao2017pyramid}
Zhao H, Shi J, Qi X, et~al (2017) Pyramid scene parsing network. In:
  Proceedings of the IEEE conference on computer vision and pattern
  recognition, pp 2881--2890

\end{thebibliography}


\end{document}